**Design and Simulation of Voltage Amplidyne System using Robust Control Technique**

Mustefa Jibril [1], Messay Tadese [2], Eliyas Alemayehu Tadese [3]

[1.] School of Electrical & Computer Engineering, Dire Dawa Institute of Technology, Dire Dawa, Ethiopia.
[2.] School of Electrical & Computer Engineering, Dire Dawa Institute of Technology, Dire Dawa, Ethiopia
[3.] Faculty of Electrical & Computer Engineering, Jimma Institute of Technology, Jimma, Ethiopia
mustefazinet1981@gmail.com

**Abstract:** In this paper, modelling designing and simulation of a simple voltage amplidyne system is done using robust control theory. In order to increase the performance of the voltage amplidyne system with H optimal control synthesis and H optimal control synthesis via-iteration controllers are used. The open loop response of the voltage amplidyne system shows that the system can amplify the input 7 times. Comparison of the voltage amplidyne system with H optimal control synthesis and H optimal control synthesis via-iteration controllers to track a desired step input have been done. Finally, the comparative simulation results prove the effectiveness of the proposed voltage amplidyne system with H optimal control synthesis controller in improving the percentage overshoot and the settling time.

## 1. Introduction

An amplidyne is an electromechanical amplifier invented prior to World War II by Ernst Alexanderson. It consists of an electric powered motor riding a DC generator. The signal to be amplified is carried out to the generator's field winding, and its output voltage is an amplified reproduction of the field current. The amplidyne is used in enterprise in excessive power servo and manage systems, to increase low power manage signals to control powerful electric powered motors. An amplidyne incorporates an electric powered motor which turns a generator on the identical shaft. Unlike an ordinary motor-generator, the cause of an amplidyne isn't always to generate a consistent voltage but to generate a voltage proportional to an Input current, to modify the input. The motor affords the power, turning the generator at a constant velocity, and the signal to be amplified is applied to the generator's field winding.

## 2. Mathematical Modelling of the system

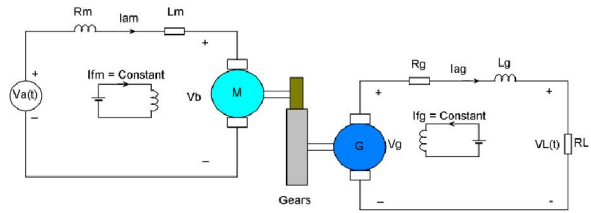

Figure 1 Voltage amplidyne system

**The Motor Side**

Assume

The stator current is constant therefore the magnetic flux is constant

$$\phi(t) = K_f I_f \qquad (1)$$

The motor torque is proportional to the armature current and the flux

$$T_m(t) = K_m i_a(t) \phi = K_1 i_a \qquad (2)$$

The voltage vb is proportional to the angular speed of the motor

$$V_b(t) = K_2 \omega_m(t) \qquad (3)$$

Applying KVL to the motor circuit

$$L_m \frac{di_a(t)}{dt} + R_m i_a(t) + V_b(t) = V_a(t) \qquad (4)$$

Substituting Equation (3) in to Equation (4) yields:

$$L_m \frac{di_a(t)}{dt} + R_m i_a(t) + K_2 \omega_m(t) = V_a(t) \qquad (5)$$

The rotational equation of the motor is

$$J_m \frac{d\omega_m(t)}{dt} + B_m \omega_m(t) = T_m(t) = K_1 i_a(t) \qquad (6)$$

Taking the Laplace Transform of Equation (5) and Equation (6)

$$I_a(s)[sL_m + R_m] + K_2 \omega_m(s) = V_a(s) \qquad (7)$$

$$\omega_m(s)[sJ_m + B_m] = K_1 I_a(s) \qquad (8)$$

Substituting Equation (8) into Equation (7) for Ia yields the transfer function between the input voltage and the output angular speed

$$\frac{\omega_m(s)}{V_a(s)} = \frac{K_1}{\left[(sJ_m + B_m)(sL_m + R_m) + K_1 K_2\right]} \qquad (9)$$

**For the generator**

The field current is constant therefore the flux is constant

The generated voltage is proportional to the angular speed, flux and field current

$$V_g = K_g \omega \phi I_{fg} \qquad (10)$$

$$\phi I_{fg} = K_f$$

$$V_g = K_g K_f \omega = K_3 \omega \qquad (11)$$

Applying KVL to the generator circuit

$$L_g \frac{di_{ag}(t)}{dt} + (R_g + R_L)i_{ag}(t) = V_g \qquad (12)$$

Substituting Equation (10) in to Equation (12) yields:

$$L_g \frac{di_{ag}(t)}{dt} + (R_g + R_L)i_{ag}(t) = K_3 \omega(t) \qquad (13)$$

Taking the Laplace Transform of Equation (13)

$$I_{ag}(s)\left[sL_g + (R_g + R_L)\right] = K_3 \omega(s) \qquad (14)$$

The transfer function between the input angular speed and the output current is

$$\frac{I_{ag}(s)}{\omega(s)} = \frac{K_3}{sL_g + (R_g + R_L)} \qquad (15)$$

Simply the load voltage is

$$V_L(s) = R_L I_{ag}(s) \qquad (16)$$

Substituting Equation (16) in to Equation (15) yields:

$$\frac{V_L(s)}{\omega(s)} = \frac{R_L K_3}{sL_g + (R_g + R_L)} \qquad (17)$$

From the gear box relation

$$\omega(s) = N\omega_m(s) \qquad (18)$$

Substituting Equation (18) in to Equation (17) yields:

$$\frac{V_L(s)}{\omega_m(s)} = \frac{NR_L K_3}{sL_g + (R_g + R_L)} \qquad (19)$$

The transfer function between the input motor voltages to the output load voltage is found by combining Equation (9) and Equation (19) which results:

$$\frac{V_L(s)}{V_a(s)} = \frac{K_1 NR_L K_3}{\left[sL_g + (R_g + R_L)\right]\left[(sJ_m + B_m)(sL_m + R_m) + K_1 K_2\right]}$$

The parameters of the system is shown in Table 1 below.

Table 1 System parameter

| No | Parameter | Symbol | Value |
|---|---|---|---|
| 1 | Motor coil inductance | $L_m$ | 18 H |
| 2 | Motor coil resistance | $R_m$ | 27 ohm |
| 3 | Moment of inertia of the motor | $J_m$ | 66 |
| 4 | Damping coefficient of the motor | $B_m$ | 28 |
| 5 | Generator Coil inductance | $L_g$ | 16 H |
| 6 | Generator coil resistance | $R_g$ | 28 ohm |
| 7 | Load resistance | $R_L$ | 100 ohm |
| 8 | Torque constant of the motor | $K_1$ | 8 |
| 9 | Voltage constant of the motor | $K_2$ | 16 |
| 10 | Voltage constant of the generator | $K_3$ | 18 |
| 11 | Gear ratio | $N$ | 55 |

The transfer function of the system becomes

$$G(s) = \frac{41.67}{s^3 + 9.9s^2 + 16.1s + 5.95}$$

And the state space representation becomes

$$\dot{x} = \begin{pmatrix} -9.9242 & -16.1380 & -5.9529 \\ 1 & 0 & 0 \\ 0 & 1 & 0 \end{pmatrix} x + \begin{pmatrix} 1 \\ 0 \\ 0 \end{pmatrix} u$$

$$y = \begin{pmatrix} 0 & 0 & 41.6667 \end{pmatrix} x$$

## 3. The Proposed Controllers Design
### 3.1　H　Optimal Control Synthesis Controller Design

**H**　optimal control synthesis solve the small-gain infinity-norm robust control problem; i.e., find a stabilizing controller F (s) for a system P (s) such that the closed-loop transfer function satisfies the infinity-norm inequality.

$$\left\| T_{y_1 u_1} \right\|_\infty \triangleq \sup \sigma_{\max} \left( T_{y_1 u_1}(j\omega) \right) < 1$$

The block diagram of the system with **H** optimal control synthesis controller is shown in Figure 2 below

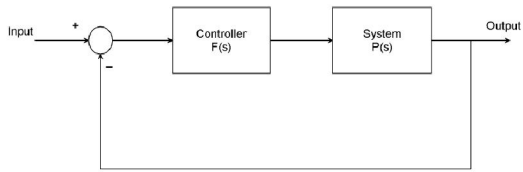

Figure 2 Block diagram of the system with **H** optimal control synthesis controller

An important use of the infinity-norm control theory is for direct shaping of closed-loop singular value Bode plots of control systems. In such cases, the system P (s) will typically be the plant augmented with suitable loop-shaping filters.

The **H**　optimal control synthesis controller transfer function is

$$F(s) = \frac{0.9956s^3 + 9.88s^2 + 16.07s + 5.926}{s^4 + 11.87s^3 + 37.39s^2 + 56.39s + 0.5602}$$

### 3.2　H　Optimal Control Synthesis via　-iteration Controller Design

**H**　optimal control synthesis via　-iteration compute the optimal **H**　controller using the loop-shifting two-Riccati formulae. The output is the optimal "　" for which the cost function can achieve under a preset tolerance.

$$\left\| \begin{bmatrix} \gamma T_{y_1 u_1} (ga \min d,;) \\ T_{y_1 u_1} (other ind,;) \end{bmatrix} \right\|_\infty \leq 1$$

The search of optimal　　stops whenever the relative error between two adjacent stable solutions is less than the tolerance specified. For most practical purposes, the tolerance can be set at 0.01 or 0.001. The block diagram of the system with **H** optimal control synthesis via　-iteration controller is shown in Figure 3 below

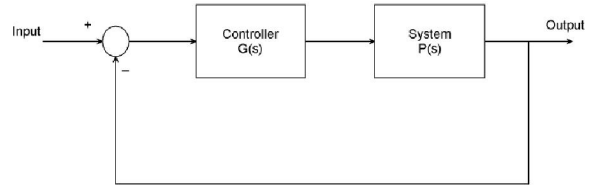

Figure 3 Block diagram of the system with **H** optimal control synthesis via　-iteration controller

The **H**　optimal control synthesis via　-iteration controller transfer function is

$$G(s) = \frac{0.5001s^3 + 4.963s^2 + 8.071s + 2.977}{s^4 + 13.39s^3 + 50.62s^2 + 62.53s + 0.6202}$$

## 4.　Result and Discussion
### 4.1　Voltage Amplidyne System Open Loop Response

The Simulink model of the open loop voltage amplidyne system and the simulation result of the system for a constant motor voltage input of 30 volt is shown in Figure 4 and Figure 5 respectively.

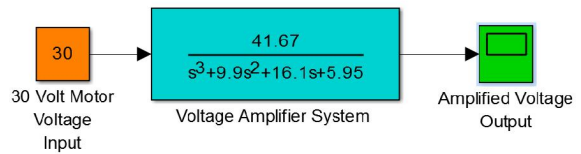

Figure 4 Simulink model of the open loop voltage amplidyne system

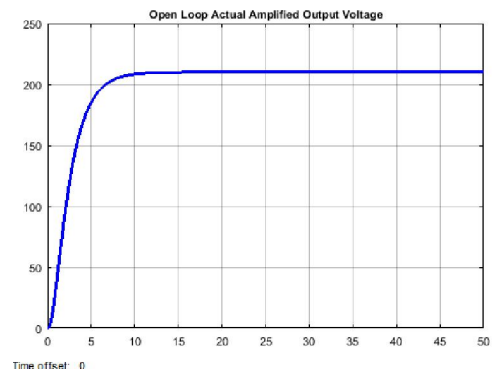

Figure 5 Simulation result

The simulation result shows that the amplidyne output voltage is 210 volt which amplifies the voltage 7 times.

**4.2    Comparison of the Proposed Controllers for Tracking a Desired Step Amplidyne Voltage**

The Simulink model of the voltage amplidyne system with $H_\infty$ optimal control synthesis and $H_\infty$ optimal control synthesis via $\mu$-iteration controllers are shown in Figure 6 below

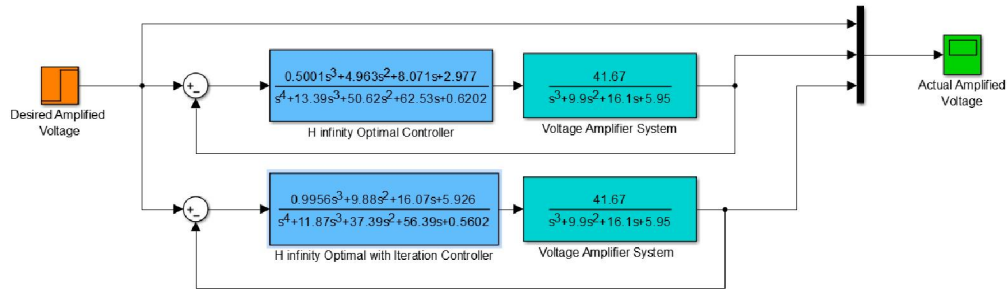

Figure 6 Simulink model of the voltage amplidyne system with $H_\infty$ optimal control synthesis and $H_\infty$ optimal control synthesis via $\mu$-iteration controllers

The simulation result of the voltage amplidyne system with $H_\infty$ optimal control synthesis and $H_\infty$ optimal control synthesis via $\mu$-iteration controllers for tracking a desired step (from 0 to 220 V) input is shown in Figure 7 below.

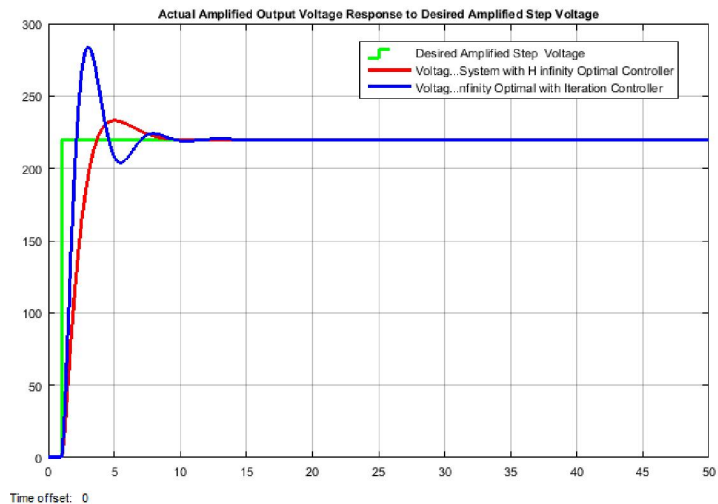

Figure 7 Simulation result

The performance data of the rise time, percentage overshoot, settling time and peak value is shown in Table 2.

Table 2 Step response data

| No | Performance Data | $H_\infty$ optimal | $H_\infty$ optimal via $\mu$-iteration |
|---|---|---|---|
| 1 | Rise time | 2.1 sec | 2 sec |
| 2 | Per. overshoot | 4.54 % | 22.72 % |
| 3 | Settling time | 9 sec | 11 sec |
| 4 | Peak value | 230    V | 270 |

**5.    Conclusion**

In this paper, a simple voltage amplidyne system is designed using a DC motor generator combination.

In order to improve the performance of the system, a robust control technique with $H_\infty$ optimal control synthesis and $H_\infty$ optimal control synthesis via $\mu$- iteration controllers are used. The open loop response of the system shows that a 7 time's amplification voltage is gained. The comparison of the proposed controllers is done to track a desired step amplified voltage and the results proves that the system with **H** optimal control synthesis controller improves the settling time and the percentage overshoot than the system with **H** optimal control synthesis via - iteration controller.

8/16/2020